# XI International Workshop on Locational Analysis and Related Problems

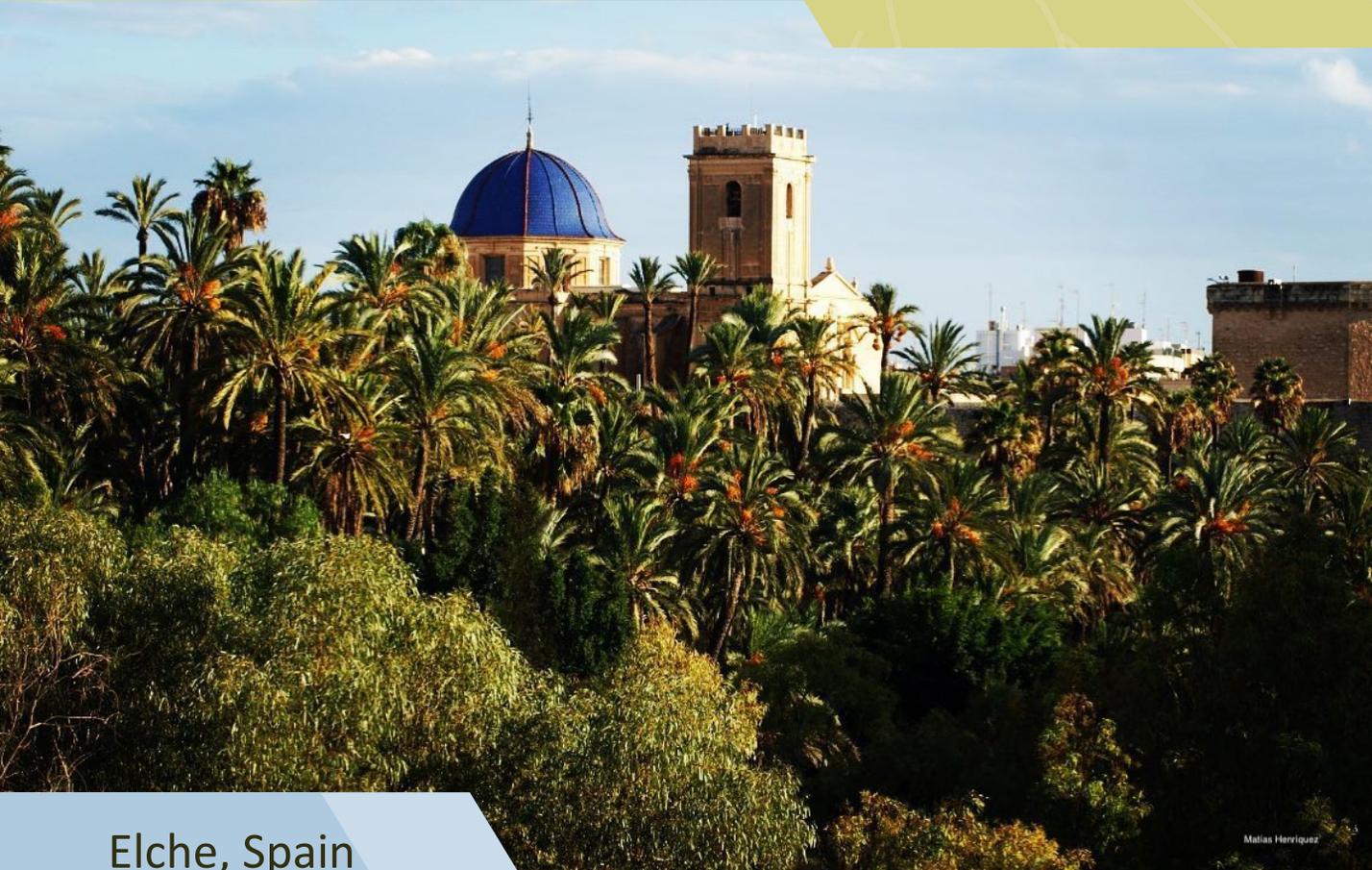

Matías Henríquez

Elche, Spain

January 31-February 1, 2022

Partners and sponsors

Red de Localización y problemas afines
Gobierno de España
Centro de Investigación Operativa  (UMH)
Prometeo 2021 (Consellería de Innovación,
            Universidades, Ciencia y Sociedad Digital)
Universitat Politècnica de Catalunya
Universidad de las Palmas de Gran Canaria
Matías Henríquez

|       | Monday Jan 31st       | Tuesday Feb 1st      |
|-------|-----------------------|----------------------|
| 9:00  | Registration          |                      |
| 9:30  | Opening Session       |                      |
| 9:45  | Session 1: Facility Location I | Session 4: Routing |
| 11:05 | Coffee break          | Coffee break         |
| 11:35 | Invited Speaker: Hande Yaman | Invited Speaker: Juan A. Mesa |
| 12:35 | Social Event: +       | Session 5: Applications II |
| 13:45 | Lunch                 | Lunch                |
| 15:30 | Session 2: Applications I | Session 6: Network design |
| 16:30 | Coffee break          | Coffee break         |
| 17:00 | Session 3: Multiperiod Problems | Session 7: Covering Problems |
| 18:00 | Network Meeting       | Social event         |
| 21:00 |                       | Conference dinner    |

PROCEEDINGS OF
THE XI INTERNATIONAL WORKSHOP
ON LOCATIONAL ANALYSIS AND
RELATED PROBLEMS (2021)

# Preface

The International Workshop on Locational Analysis and Related Problems will take place during January 31–February 1, 2022 in Elche (Spain). It is organized by the Spanish Location Network and the Location Group GE-LOCA from the Spanish Society of Statistics and Operations Research(SEIO). The Spanish Location Network is a group of more than 140 researchers from several Spanish universities organized into 7 thematic groups. The Network has been funded by the Spanish Government since 2003. This edition of the conference is organized in collaboration with project PROM-ETEO/2021/063 funded by the Valencian government.

One of the main activities of the Network is a yearly meeting aimed at promoting the communication among its members and between them and other researchers, and to contribute to the development of the location field and related problems. The last meetings have taken place in Sevilla (January 23–24, 2020), Cádiz (January 20–February 1, 2019), Segovia (September 27–29, 2017), Málaga (September 14–16, 2016), Barcelona (November 25–28, 2015), Sevilla (October 1–3, 2014), Torremolinos (Málaga, June 19–21, 2013), Granada (May 10–12, 2012), Las Palmas de Gran Canaria (February 2–5, 2011) and Sevilla (February 1–3, 2010).

The topics of interest are location analysis and related problems. This includes location models, networks, transportation, logistics, exact and heuristic solution methods, and computational geometry, among others.

The organizing committee.

## Scientific committee:

- Maria Albareda Sambola (Universitat Politécnica de Cataluña, Spain)
- Giuseppe Bruno (Università degli Studi di Napoli Federico II, Italy)
- Sergio García (University of Edinburgh, United Kingdom)
- Jörg Kalcsics (University of Edinburgh, United Kingdom)
- Alfredo Marín (Universidad de Murcia, Spain)
- Blas Pelegrín (Universidad de Murcia, Spain)
- Justo Puerto (Universidad de Sevilla, Spain)
- Antonio M. Rodríguez-Chía (Universidad de Cádiz, Spain)
- Francisco Saldanha da Gama (Universidade de Lisboa, Portugal)

## Organizing committee:

- Maria Albareda Sambola (Universitat Politècnica de Catalunya)
- Javier Alcaraz Soria (Universidad Miguel Hernández)
- Laura Antón Sánchez (Universidad Miguel Hernández)
- Marta Baldomero Naranjo (Universidad de Cádiz)
- Mercedes Landete Ruiz (Universidad Miguel Hernández)
- Marina Leal Palazón (Universidad Miguel Hernández)
- Juan Francisco Monge Ivars (Universidad Miguel Hernández)
- Alejandro Moya Martínez (Universidad Miguel Hernández)
- Juan Manuel Muñoz Ocaña (Universidad de Cádiz)
- Jessica Rodríguez Pereira (Universitat Pompeu Fabra)
- Dolores Rosa Santos Peñate (Universidad de Las Palmas de Gran Canaria)
- José Luis Sainz-Pardo Auñón (Universidad Miguel Hernández)

# Contents

PROGRAM

# Monday January 31st

## 09:00-09:30 Registration

## 09:30-09:45 Opening Session

## 09:45-11:05 Session 1: Facility Location

Upgrading Strategies in the p-Center Location Problem
*L. Anton-Sanchez, M. Landete, F. Saldanha-da-Gama*

Capacitated Close Enough Facility Location
*A. Moya-Martínez, M. Landete, J.F. Monge, S. García*

Location, Regions and Preferences
*V. Blanco, R. Gázquez, M. Leal*

Formulations for the Capacitated Dispersion Problem
*M. Landete, J. Peiró, H. Yaman*

## 11:05-11:35 Coffee break

## 11:35-12:35 Invited Speaker: Hande Yaman

Robust Alternative Fuel Refueling Station Location Problem with Routing under Decision-Dependent Flow Uncertainty

## 12:35-15:30 Social Event and Lunch

## 15:30-16:30 Session 2: Applications I

Emergency Vehicles Location: the importance of including the dispatching problem.
*J. Nelas, J. Dias*

Optimizing COVID-19 Test and Vaccine distributions
*J.L. Sainz-Pardo, J. Valero*

Locating a rectangle in the sky to get the best observation
*J.J. Salazar-González*

**16:30-17:00 Coffee break**

## 17:00-18:00 Session 3: Multiperiod Problems

An exact method for the two-stage multiperiod vehicle routing problem with depot location
*I. Gjeroska, S. García*

Multistage multiscale facility location and expansion under uncertainty
*L.F. Escudero, J.F. Monge*

How to invest to expand a firm: a new model and resolution methods
*J. Fernández, B.G.-Tóth, L. Anton-Sanchez*

## 18:00 Network meeting

# Tuesday February 1st

## 09:45-11:05 Session 4: Routing

Selective collection routes of urban solid waste by means of multi-compartment vehicles
*R. Piedra-de-la-Cuadra, J.A. Mesa, F. A. Ortega, G. Marseglia*

Multi-Depot VRP with Vehicle Interchanges: Heuristic solution
*V. Rebillas-Loredo, M. Albareda Sambola, J.A. Díaz, D.E. luna*

A new heuristic for the Driver and Vehicle Routing Problem
*B. Domínguez-Martín, I. Rodríguez-Martín, J.J. Salazar-González*

## 11:05-11:35 Coffee break

## 11:35-12:35 Invited Speaker: Juan A. Mesa

Pair-demand Covering Facility Location and Network Design Problems

## 12:35-13:35 Session 5: Applications II

A column-and-row generation algorithm for allocating airport slots
*P. Fermín Cueto, S. García, M. F. Anjos*

A locational analysis perspective of deregulation policies in the pharmaceutical sector
*G. Bruno, M. Cavola, A. Diglio, J. Elizalde, C. Piccolo*

The discrete ordered median problem for clustering STEM-image intensities
*J.J Calvino, M. López-Haro, J.M. Muñoz-Ocaña, A.M. Rodríguez-Chía*

## 13:45-15:30 Lunch

## 15:30-16:30 Session 6: Netwok Design

An Iterated Greedy Matheuristic for Solving the Stochastic Railway Network Construction Scheduling Problem
*D. Canca, G. Laporte*

Multiple Allocation P-Hub Location Problem explicitly considering Users' preferences
*N. Zerega, A. Lüer-Villagra*

Profit-maximizing hub network design under hub congestion and time-sensitive demands
*C.A. Domínguez, E. Fernández, A. Lüer-Villagra*

## 16:30-17:00 Coffee break

## 17:00-18:00 Session 7: Covering Problems

On the complexity of the upgrading version of the Maximal Covering Location Problem
*M. Baldomero-Naranjo, J. Kalcsics, A. M. Rodríguez-Chía*

Fairness in Maximal Covering Facility Location Problems
*V. Blanco, R. Gázquez*

Hybridizing discrete and continuous maximal covering location problems
*V. Blanco, R. Gázquez, F. Saldanha-da-Gama*

## 18:00 Social Event

## 21:00 Conference dinner

INVITED SPEAKERS

# Robust Alternative Fuel Refueling Station Location Problem with Routing under Decision-Dependent Flow Uncertainty[*]

Özlem Mahmutoğulları[1] and Hande Yaman[2]

[1]*ORSTAT, FEB, KU Leuven, 3000 Leuven, Belgium*   ozlem.mahmutoullar@kuleuven.be

[2]*ORSTAT, FEB, KU Leuven, 3000 Leuven, Belgium*   hande.yaman@kuleuven.be

## 1.     Introduction

Transportation is heavily dependent on fossil fuels, especially petroleum-based products. Using alternative fuel vehicles is a solution to break the transportation sector's reliance on consuming fossil fuels. The lack of alternative fuel station (AFS) infrastructure and the rather limited range of alternative fuel vehicles (AFVs) are two significant obstacles that are slowing down the introduction of AFVs. In this regard, the refueling station location problem (RSLP) has recently started to be studied in the literature. In the RSLP, the AFSs are located on the drivers' predetermined paths. Since the drivers may sometimes tolerate deviating from their paths to refuel their vehicles, the RSLP with routing (RSLP-R) extends the RSLP and determines the locations of stations and routes of drivers simultaneously.

It is likely to observe uncertainties in the flows because the rollout of AFVs and the development of the AFS network are still at their initial stages. Moreover, the statistical data shows that the number of AFSs has a significant impact on the number of AFVs. It is thus important to consider that the availability of AFSs in the neighborhood affects the proliferation of AFVs during the development of infrastructure. Hence, we incorporate robustness and decision-dependency into the RSLP-R.

[*]Research supported the KU Leuven grant 3H180528

## 2.     Problem and Solution Methods

The RSLP-R is defined on a road network and aims to maximize the total amount of AFV flows that can be refueled by locating a predetermined number of AFSs on the network by considering the willingness of drivers to deviate from their shortest paths to refuel their vehicles as well as the limited range of the vehicles. We use the deterministic problem introduced by [1] and introduce our flow uncertainty set using the hybrid model ([2]). The hybrid model comprises a hose model and an interval model. We define the hybrid uncertainty set of the vehicle flows under the impact of station location decisions. We suppose that, when a new station is opened, vehicle flows in the neighborhood increase. We derive two mathematical programming formulations. As the problem size grows, we encounter difficulties in solving these models, and thus we propose a Benders reformulation. We solve this formulation using a branch-and-cut algorithm. The separation, which is exact and polynomial, is done by inspection.

## 3.     Computational Results

We use four different sized data sets to perform our computational experiments. The first one is a commonly used data set in the RSLP literature. We generated the other data sets based on the road network of Belgium. We perform the following computational experiments: We first compare the performances of the proposed solution methods. We observe that the Benders reformulation outperforms the other formulations. Then, we investigate the changes in station locations and total covered flows when the optimal solutions of the deterministic, robust (without decision-dependency), and decision-dependent robust problems are employed. We also analyze the changes under different parameter settings. We observe that recognizing the uncertainty in flows and the decision-dependency of uncertain flow realizations may lead to significant gains in the total AFV flows covered.

# Pair-demand Covering Facility Location and Network Design Problems[*]

Juan A. Mesa [1]

[1]*Departamento de Matemática Aplicada II, Universidad de Sevilla, Sevilla, Spain*
jmesa@us.es

Covering along with median and center are three classical branches of Facility Location Problems. Covering problems have been extended to extensive facility location, where facilities are too large to be represented as isolated points, as well as to network design where a (sub)network is to be selected from an (physical or not) underlying network with the aim of being used by traffic of goods or people.

Covering problems in graphs have attracted the attention of researchers since the middle of last century. As far as the author is aware the first papers on the vertex-covering problem were due to Berge (1957 [1], and Norman and Rabin (1959) [2]. The problem dealt with in these papers was to find a subset of vertices in a graph with minimum caridinality such that each edge is incident to at least one vertex.This problem is related with the set-covering problem is which a family of sets is given and the minimal subfamily which union contains all the element is sought. The decision versions of both problems were proved to be NP-complete (Karp, 1972 [3]). The vertex-covering problem was formulated as a integer linear programming model and solved by using Boolen functions by Hakimi (1965, [4]). This problem was applied to the location of emergency services by Toregas et al. (1971, [5]). They assume a vertex is covered by other if it is within a given coverage distance from the other. When the number of facilities to be located is fixed then the maximal covering location problem arise (Church and ReVelle, 1974, [6]) in which problem each vertex has an associated population and the objective is to cover the maximum population within a given distance threshold. Since then many variants and extensions of the vertex-covering and maximal covering problems have been researched (see Schilling el at., 1993 [7], Farahani et al. 2012 [8], Church and Murray, 2018 [9] and García and Marín, 2019 [10] for a review on models, theoretical results and solving procedures regarding point covering problems.)

---

[*] Research partially supported by the Spanish Ministry of Science and Innovation through project RED2018-102363-T, Ministerio de Investigación (Spain)/FEDER under grant PID2019-533 106205GB-I00, and Operational Programme FEDER-Andalucía under grant US-1381656.

An extensive facility location problem on networks consists of locating a subbgraph such that optimizes some objective function with some constraints regarding demand points (Puerto et al. 2018, [11], Mesa 2018, [12]). The extensive vertex-covering and maximal covering problems are extensions of the point vertex-covering and maximal covering problems where the number of vertices of the coverage is substituted by the length or other characteristic of the covering subgraph.

All the previously reviewed research considers origin-facility or facility-destination systems, where the demand is satisfied by accessing the point or the extensive facilities. However, in many cases to reach or be closed the facility is not enough for completing the service. In many telecommunication, transportation, public services and other systems the demand uses the facility as an intermediate instead of a final destination. In these cases the demand is given by pairs instead of single points, and usually each pair has an associated weight indicating the traffic between the origin and the destination. This is for example the case of public transportation networks, where a customer have to spend a time to reach the stop/station from its origin, then wait for the next service, spend an in-vehicle time, and finally reach its destination. When planning a new network, often there exists a network already functioning and offering its service to the same origin-destination pairs. For example, in order to improve the mobility of a big city or metropolitan area, a new rapid transit system is planned. The current transit system could be more dense than the rapid transit planned but is slower since uses the same right-of-way than the private traffic. Thus, in some way both systems compete between then and both compete with the private mode of transportation. In a recent paper [13] Benders decomposition is applied to the Maximal Covering Network Design and the Partial Covering Network Design problems.

A review of origin-facility-destination demand Network Design problems is included in the paper by Contreras and Fernández, 2015 [14], in which mathematical programming models for Location-Network Design problems are formulated. Within this framework, Hub Location and Network Design problems have been intensively researched. Integer programming formulations for the Hub Set Covering Location and the Hub Max-imal Covering problems were provided in the paper by Campbell, 1995 [15]. For more details on Hub Location problems we refer the reader to the chapter (2019), [16].

Most of the research on origin-facility-destination covering problems adopts a discrete formulations. However, some applications have a continuous nature, either because the facilities can be located on edges of a given network or because the space for locating the networks is a continuous one. In these cases a continuous formulation better fits to the problem. Nevertheless, the research on continuous pair covering Location and Network Design is scarce. An example of these problems can be found in the papers by López-de-los-Mozos et al. 2017 [17], and López-de-los-Mozos and Mesa [18], in which new transfer points in a network are located for satisfying the pair demand. A review on the location of dimensional facilities in continuous spaces can be found in Schoebel, 2019 [19].

Therefore, when the demand is given by pairs of points instead of single points, new covering Location and Network Design problem arise. In this tallk, we will revise some problems that have been already researched and their solving approaches, propose a classification and a framework for these problems, and suggest some lacks in the state-of-the-art.

ABSTRACTS

# Upgrading Strategies in the $p$-Center Location Problem [*]

Laura Anton-Sanchez,[1] Mercedes Landete,[1] and Francisco Saldanha-da-Gama[2]

[1]*Departamento de Estadística, Matemáticas e Informática, Centro de Investigación Operativa, Universidad Miguel Hernández, Spain,*   l.anton@umh.es        landete@umh.es

[2]*Departamento de Estatística e Investigação Operacional, Centro de Matemática, Aplicações Fundamentais e Investigação Operacional, Faculdade de Ciências, Universidade de Lisboa, Portugal,*   fsgama@ciencias.ulisboa.pt

## 1.     Introduction

Given a set of nodes in a metric space, the $p$-center problem consists of determining at most $p$ points in such a way that the maximum distance between the given nodes and the closest centers is minimized [2].

The $p$-center problem and its variants have many applications among which we can point out those in Telecommunications, Logistics, Emergency facility location, etc. To the best of the authors' knowledge, the literature on the $p$-center problems assumes that the costs (distances, travel times, etc) are known beforehand and do not change. Nevertheless, in practice, one may ask whether a better solution can be achieved by somehow compressing/reducing the allocation costs thus obtaining the so-called upgraded solutions. In this work we investigate different upgrading strategies in the context of the $p$-center problem.

---

[*]This work was partially supported by the grants PID2019-105952GB-I00/ AEI /10.13039/ 501100011033 and PGC2018-099428-B-100 by the Spanish Ministry of Science and Innovation, PROMETEO/2021/063 by the governments of Spain and the Valencian Community, and UIDB/04561/2020 by National Funding from FCT — Fundação para a Ciência e Tecnologia, Portugal.

# 2. Upgrading connections and facilities

We focus on the so-called *unweighted vertex-restricted $p$-center problem*. We consider the possibility of upgrading a set of connections to different facilities, as well as the possibility of upgrading entire facilities, i.e., upgrading all connections to an open facility. Further, we consider two perspectives when it comes to upgrading decisions (connections or facilities): (i) a limit is imposed on the number of connections or facilities that can be upgraded; (ii) a budget exists that limits the upgrades that can be made. For other upgrading versions of location problems, see e.g. [1, 5].

We introduce different MILP models for the different upgrading strategies. Our models are based on those previously proposed by [3] and [4]. Furthermore, we derive lower and upper bounds for the new models. We show that a significant decrease in the optimal covering cost can be attained by upgrading connections or facilities. Therefore, the information provided by the new models can be extremely useful to a decision-maker because together with the location decision, the models directly seek to find structures underlying the problem that can be "upgraded" in such a way that a better after-upgrading solution is obtained. Moreover, the research done in this work indicate different directions for future work in the topic.

# On the complexity of the upgrading version of the Maximal Covering Location Problem[*]

Marta Baldomero-Naranjo,[1] Jörg Kalcsics,[2] and Antonio M. Rodríguez-Chía[1]

[1]*Departamento de Estadística e Investigación Operativa, Universidad de Cádiz, Spain,* marta.baldomero@uca.es antonio.rodriguezchia@uca.es

[2]*School of Mathematics, University of Edinburgh, UK,*  joerg.kalcsics@ed.ac.uk

We study the upgrading version of the maximal covering location problem with edge length modifications on networks (Up-MCLP). The Up-MCLP aims at locating $p$ facilities to maximize the coverage taking into account that the length of the edges can be reduced subject to a budget constraint. Therefore, we look for both solutions: the optimal location of $p$ facilities and the optimal upgraded network. Note that for each edge, we are given its current length, an upper bound on the maximal reduction of its length, and a cost per unit of reduction (which can be different for each edge). Furthermore, a total budget for reductions is given.

In this talk, we focus on the complexity of this problem. Since it is a particular case of the Maximal Covering Location Problem, the Up-MCLP is NP-hard on general graphs. We analyze different types of graphs (paths, trees, stars, etc.) and study the complexity of the single-facility and the multi-facility version of the problem under different assumptions on the model parameters. We prove that this problem can be solved in polynomial time and pseudo-polynomial time in some particular cases. We derive algorithms for solving them. Moreover, we show several particular cases in which the problem is NP-hard.

[*]Research partially supported by the Spanish Ministry of Science and Innovation through project RED2018-102363-T

# Fairness in Maximal Covering Facility Location Problems

Víctor Blanco,[1] and Ricardo Gázquez[1]

[1] *Institute of Mathematics (IMAG), Universidad de Granada, Granada, Spain,* vblanco@ugr.es rgazquez@ugr.es

According to the Cambridge Dictionary the term *fairness* is defined as "the quality of treating people equally or in a way that is right or reasonable". It is an abstract but widely studied concept in Decision Sciences in which some type of indivisible resources are to be shared among different agents. Fair solutions should imply impartiality, justice and equity allocation patterns, which are usually quantified by means of inequality measures that are usually minimized. The importance of fairness issues in resource allocation problems has been recognized and well studied in a variety of settings with tons of applications in different fields.

The covering location problem is a core problem in Location Science (see [3]). Here, we focus in Maximal Covering Location Problem (MCLP) in which it is assumed the existence of a budget for opening facilities and the goal is to accommodate it to satisfy as much demand of the users as possible. This problem has attracted the attention of many researchers since its introduction by Church and ReVelle [2], both because its practical interest in different disciplines (see [4]) and the mathematical challenges it poses. The efficiency measure used in the MCLP is the overall covered demand, that is, as much covered demand the better. However, when one looks at the individual utilities of each of the constructed facilities, one may obtain solutions with highly saturated facilities in contrast to others that only cover a small amount of demand, which results in an unfair system from the facilities' perspective.

In this paper we provide a general mathematical programming based framework to incorporate fairness measures from the facilities' perspective to Discrete and Continuous MCLPs. In a fairly ideal solution, one would desire to "independently" maximize the covered demand of each of the services, not affecting negatively to the coverage of the others. However, since the demands are usually indivisible, in most cases, an advantageous solution for one service harms others. As already happens in other decision problems, one may prefer to slightly sacrifice the overall covered demand in order to equalize the different covered demands among the open services. This might be the case of the location of public schools, in which it is preferable to find an homogeneous distribution of kids among the schools, or the location of routers with high capacities, where a "good" location for them would be the one in which the performance of all the routers can be better used instead of saturating some and leaving others more free. We incorporate fairness criteria into covering location models by means of two powerful tools: (1) the Ordered Weighted Averaging (OWA) aggregation operators of the covered demands of each service (introduced by Yager [5]), and (2) the $\alpha$-fairness scheme that, depending on the value of the parameter $\alpha$, may represent classic measures of fairness (introduced by Atkinson [1]).

# Location, Regions and Preferences [*]

Víctor Blanco,[1] Ricardo Gázquez,[1] and Marina Leal[2]

[1]*Institute of Mathematics, Universidad de Granada*

[2]*Operational Research Center (CIO), Universidad Miguel Hernández*

Customer preferences when purchasing goods and services have been widely analyzed by Utility Theory in Economics. This theory makes the decisions about the best ways of satisfying customer's demands easier for the companies [1, 6]. In this work, we analyze the incorporation of preference measures to the continuous facility location problem with regional demands.

In continuous location problems with regional demands, it has been analyzed the minimization of the expected distance from the new facility to the demand region [2], the expected demand [7] or the maximum distance between the new facility and the region [4, 5] among others. To the best of our knowledge, there has not been attempt to incorporate user's preferences in these type of location problems.

We propose a framework to deal with the incorporation of customer's preferences in continuous location problems with demand regions. We assume that each user is served in a spatial region and that a preference function is given over each region. The service points of the demand regions are served from a new *central* facility. The goal is to determine the locations of the service points in each demand region and the location of the new central facility at minimum transportation cost and reaching certain preference level of the customers. We consider different preference functions (see Figure 1 for an example with linear preferences functions). An application of the proposed model can be found, for instance, in the location design of central storehouses and containers of e-commerce companies.

---

[*]Research partially supported by the Spanish Ministry of Science and Innovation through project PID2020-114594GB-C21, also by project Junta de Andalucía P18-FR-1422 and Project I+D+i FEDER Andalucía US-1256951

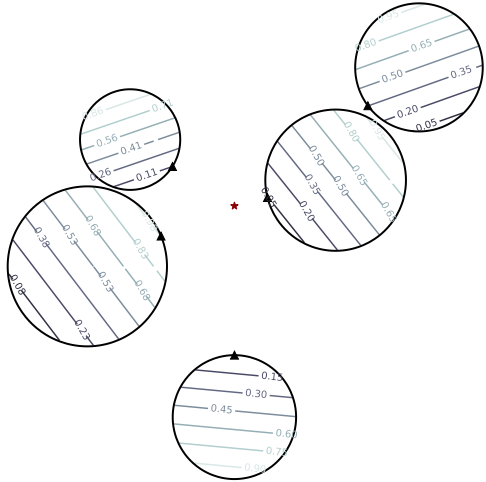

*Figure 1.* Illustration of a Continuous Location Problem with Demand Regions and Linear Preference Funtions (the star represents the central facility and the triangles the service points in each demand region).

# Hybridizing discrete and continuous maximal covering location problems[*]

Víctor Blanco,[1] Ricardo Gázquez,[1] and Francisco Saldanha-da-Gama[2]

[1]*Institute of Mathematics (IMAG), Universidad de Granada, Spain.* {vblanco, rgazquez}@ugr.es

[2]*Centro de Matemática, Aplicações Fundamentais e Investigação Operacional, Universidade de Lisboa, Portugal.* faconceicao@fc.ul.pt

## 1.     Introduction

Among the most popular problems in Location Analysis are those in which a user can receive a service in case it is located close enough to an open facility providing it. These problems are usually called *Covering Location problems*. In case it is assumed the existence of a budget for opening facilities and the goal is to accommodate it to satisfy as much demand of the users as possible, the problems belong to the family of *Maximal Covering Location Problems* (MCLP) that have attracted the attention of many researchers since its introduction in [2], both because its practical interest in different disciplines and the mathematical challenges it poses. The interested readers are referred to the surveys in [1] and [3] for further details on covering location problems.

A feature shared by most of the existing literature focusing the MCLP concerns the existence of a single type of facility. However, in practice, this may not be the case. If not by other reasons, the progressive technology development often calls for older equipment that is still operational to be used together with more recent one. Another possibility emerges when two technologies can be looked at as complementing each other. For instance, when locating equipment for early fire detection in forests, surveillance facilities requiring human resources operating them may be complemented

[*]Research partially supported by the Spanish Ministry of Science and Innovation through project RED2018-102363-T

with equipment such as remotely controlled cameras to ensure a better coverage of the area of interest. When facilities can be installed in different phases (e.g. multi-period facility location) the facilities to be located in each phase can also be looked at as belonging to a different group (that we still call type) of facilities.

We investigate here the MCLP with multiple facility types. We assume that the number of facilities of each type to be located is known beforehand and that each type of facilities is characterized by the shape of their coverage areas and the metric space from which they are selected. By considering a hybrid setting it becomes possible to take advantage from choosing some services in finite sets of pre-specified preferred locations and then deciding flexible positions of the servers in the whole space. This setting can be useful, for instance, in telecommunications networks with a certain number of the servers being located inside adequately prepared infrastructures and additional servers being located at any place in the given space. The goal is of course to capture as much demand as possible no matter the equipment doing it. The continuous facilities can be looked at as a set of servers to be located in the future and that must complement the equipment located in a discrete setting.

First, we present a general mathematical programming formulation for the problem. Afterwards, motivated by some practical settings we investigate the hybridization of discrete and continuous facility location. We consider that several types of facilities are to be selected in finite sets of possibilities (one for each type) whereas the other types of facilities can be located continuously in the whole space. Finally, we report the results of an extensive battery of computational experiments performed to assess the methodological contribution of this work. The data consists of up to 920 demand nodes using real geographical and demographic data.

# A locational analysis perspective of deregulation policies in the pharmaceutical sector

Giuseppe Bruno,[1] Manuel Cavola,[1], Antonio Diglio[1], Javier Elizalde[2] and Carmela Piccolo[1]

[1]*Università degli Studi di Napoli Federico II, Department of Industrial Engineering (DII), Piazzale Tecchio, 80 - 80125 Naples, Italy,*   {giuseppe.bruno, manuel.cavola, antonio.diglio, carmela.piccolo}@unina.it

[2]*Facultad de Ciencias Económicas y Empresariales, Universidad de Navarra, Campus Universitario, 31080 Pamplona, Spain,*   jelizalde@unav.es

The retail pharmaceutical sector has been characterized in the last decades by profound deregulation policies, aiming at fostering the entry process of new competitors in the market to improve users' access and quality of services. Extending a previous study by the authors ([1]), the present work explores the effects of such reforms in the Navarre Region (Spain), where the existing entry restrictions were relaxed in 2000. Using a GIS-based methodology and very disaggregated real data, we show that the massive entry of new pharmacies improved users' access in urban and rural areas but led to intense agglomeration of competitors within limited distances, thus failing to result in a more fair spatial competition.

We then propose solving a mathematical programming model that can support policy-makers for an ex-ante assessment of the (spatial) effects determined by changing regulatory frameworks. We cast the model as the facility location problem with threshold requirements introduced by [2], which is further enhanced in this work with explicit dispersion criteria among facilities.

Specifically, given the presence of an existing set of already located drugstores $J_\epsilon$, we assume that a decision-maker is interested in regulating the entry process of new drugstores across a set of potential locations $J_n$. The model takes into account different aspects of practical relevance in pharmaceutical market regulation. In particular, we assume that at most $F$ phar-macies can coexist in the market (*demographic criterion*) and that new entrants should locate at least at a distance $\bar{c}$ from another (old or new) competitor (*dispersion criteria*). The objective is to minimize the distance users must travel to reach the closest active drugstore, i.e. the *average accessibility* distance. Such an objective fosters the location of the highest possible number of pharmacies, which may be suggested to locate in peripheral areas with scarce customers. In order to provide an entry incentive, we impose that a minimum number of users is assigned to each new entrant. In other words, we are granting a minimum market share ($MS$) to newly located pharmacies. This way, the model pursues the best users' accessibility conditions while: (i) protecting - through spatial dispersion criteria - extant competitors from the threat posed by new entrants; (ii) stimulating new entries with minimum (guaranteed) quotas of customers.

Results from our computational experiments on the entire Navarre Region demonstrate that, if adequately oriented at guaranteeing predefined minimum profitability to new entrants, spatial deregulation policies may result in win-win scenarios for both users and pharmacies. Indeed, the obtained solutions (almost) perfectly replicate the current accessibility conditions while ensuring more equitable market shares' distribution.

The present work intends to contribute to the ongoing debate on the effects of policy reforms in the pharmaceutical sector and, through an indepth spatial analysis and a computational experience on a real case study, it aims at showing the benefit of these analytical tools to inform and improve the practice of decision-making in public services ([3]) or, as per the case at hand, in regulated markets and sectors of general economic interest.

# The dicrete ordered median problem for clustering STEM-image intensities [*]

José J. Calvino[3], Miguel López-Haro[3], Juan M. Muñoz-Ocaña[1], Justo Puerto[2] and Aantonio M. Rodríguez-Chía [1]

[1]*Departamento de Estadística e Investigación Operativa, Universidad de Cádiz, Cádiz, Spain,*  juanmanuel.munoz@uca.es, antonio.rodriguezchia@uca.es

[2]*IMUS, Universidad de Sevilla, Sevilla, Spain,*  puerto@us.es

[3]*Departamento de Ciencia de los Materiales e Ingeniería Metalúrgica y Química Inorgánica, Universidad de Cádiz, Cádiz, Spain*   miguel.lopezharo@uca.es, jose.calvino@uca.es

Electron tomography is a technique for imaging three-dimensional structures of materials at nanometer scale. This technique consists on reconstructing nano-objects thanks to projections provided by a electron microscope from different tilt angles. The Scanning-Transmission Electron Microscope images obtained are used for identifying the elements that constitute the nano-objects under study. This recognition procedure is known as segmentation which consists of classifying the image intensities into different clusters.

Classical segmentation models stand out for their ability to provide one segmentation of the original image very quickly and with low computational burden [1]. However, they do not usually achieve high quality segmentations with a small number of clusters to classify the different elements which compose the structures represented in the image.

The main idea behind this work is to apply the ordered median problem to locate $p$ intensities as the representatives of $p$ different clusters and allocate every intensity to one cluster representative [2]. The advantage of using this function is its good adaptability to the different types of particles to be studied due to the wide range of vector weights that can be cast [3].

[*]Research partially supported by the Spanish Ministry of Science and Innovation through project RED2018-102363-T

Moreover, to reduce the computational time needed to solve these problems, some improvements are introduced for the formulations by taking advantage of the vector weight structure. These alternative improvements are based on the idea developed in [4]. Finally, we propose different ways of analysing the quality of the segmentations provided by our approach using different choices of the vector weights in some real instances.

# An Iterated Greedy Matheuristic for Solving the Stochastic Railway Network Construction Scheduling Problem

David Canca [1] and Gilbert Laporte [2]

[1] *Department of Industrial Engineering and Management Sciecnce, School of Engineering, Av. de los Descubrimientos s/n, 41092, Seville, Spain,*  dco@us.es

[2] *HEC Montréal, CIRRELT 3000 Chemin de la Côte-Sainte-Catherine, Montréal, QC H3T 2A7, Montréal Canada,*  gilbert.laporte@cirrelt.net

We propose an iterated greedy matheuristic for efficiently solving stochastic railway rapid transit transportation network construction scheduling problems, where both the construction duration of the segments and the passenger demand rate of increase are considered stochastic. The problem consists of sequencing the construction of lines of a urban transportation network with the aim of maximizing the discounted long-term profit . This problem can be viewed as a resource-constrained project scheduling problem, where both the budget and the available construction equipment act as resources. We consider that partial lines can be put into operation, thus benefiting users with a partial and quick usage of the network infrastructure. In this situation, both the costs and the revenues dependent on the schedule. After analyzing some characteristics of the best solutions, we propose an iterated greedy matheuristic for solving the real-size network construction scheduling problems. To illustrate our methodology we apply the algorithm to the construction of the full metro network of the city of Seville.

# Profit-maximizing hub network design under hub congestion and time-sensitive demands[*]

Carmen-Ana Domínguez,[1] Elena Fernández,[1] and Armin Lüer-Villagra[2]

[1]*Department of Statistics and Operational Research, Universidad de Cádiz, Puerto Real, Spain,*　carmenana.dominguez@uca.es, elena.fernandez@uca.es

[2]*Department of Engineering Sciences, Universidad Andres Bello, Antonio Varas 880, Santiago, Chile,*　armin.luer@unab.cl

Hub location models are used to design transportation networks for airlines, parcel delivery, LTL truck companies, etc. There are previous works considering congestion, time-sensitive demands, and profit maximization, but not necessarily at the same time. We formulate and solve a profit-maximizing hub network design problem considering simultaneously hub congestion and time-sensitive demands through stepwise functions. The resulting formulation is very challenging to solve up to optimality, resulting in large optimality gaps. We develop variable fixing procedures as well as some families of valid inequalities. Preliminary results are encouraging.

## Introduction

Hub location is an active research area, as shown by the frequent literature reviews [2, 4]. Current topics of interest include extensions of earlier models like, for instance, incorporating capacity selection and/or congestion [3, 5, 6], explicitly considering sensitive demands [7], or moving from cost minimization to profit maximization [1, 8].

　　To the best of our knowledge, no previous studies jointly consider the above three modeling aspects. Our contribution is to formulate and solve a profit-maximizing hub network design problem that incorporates simultaneously hub congestion, time-sensitive demands through step-wise functions, where service paths with one or two stops are allowed.

[*]Research partially supported by the Spanish Ministry of Science and Innovation through project MINECO MTM2019-105824GB-I00 and RED2018-102363-T, and by the Chilean FONDECYT through grant 1200706.

A stepwise function on transportation times is used to model demand. Hub congestion is expressed in terms of processing times at the hubs, which are also modeled as a stepwise function. A profit is obtained from captured demand, while costs include fixed setup cost for enabling hubs and inter-hub edges, as well as the usual transportation costs.

We develop a set of conditions to either fix variables or add additional constraints to the formulation. Families of valid inequalities are also presented together with their separation procedures. These allow us to improve the LP bound of our formulation, decreasing the computational time required.

# A new heuristic for the Driver and Vehicle Routing Problem*

Bencomo Domínguez-Martín,[1] Inmaculada Rodríguez-Martín,[1] and Juan-José Salazar-González[1]

[1]*Department of Mathematics, Statistics, and Operations Research, University of La Laguna, Tenerife, Spain,*  bdomingu@ull.edu.es, jjsalaza@ull.edu.es, irguez@ull.edu.es

The *Driver and Vehicle Routing Problem* (DVRP) addressed here was introduced by Domínguez-Martín et al. [1] ant it is defined as follows. We are given two depots, where a given number of vehicles and drivers are based, and a set of customers. Each customer must be served by a vehicle and a driver. Vehicles start their routes at their base depot and end at the other depot, while drivers must start and end their routes at their base depot. The vehicles have to be always led by a driver, and drivers need a vehicle to move from one location to another, either driving themselves or traveling as passengers. When there are more than one driver in a vehicle, any of them can lead the vehicle. The duration of a driver route is the time between the departure from and the arrival to the depot, and it includes the time driving and traveling as passenger. Moreover, drivers' routes cannot exceed a given time duration. Drivers can switch vehicles only at some given points known as exchange locations, which are the only customer locations that can be visited by more than one vehicle. The objective is to design the routes of the vehicles and the drivers in order to minimize the total drivers' routes cost.

We present a new heuristic method for the DVRP that provides high quality solutions for the instances considered. The first phase of the algorithm creates driver's routes using a constructive method, and the second phase improves those routes through local search. The two phases are embedded in a multistart loop. Vehicles' routes can be derived from the

*Research partially supported by the Spanish Ministry of Science and Innovation through project PID2019-104928RB-I00

drivers'routes to end up with a feasible DVRP solution. This heuristic provides good solutions for the benchmark instances in the literature, and is able to cope with instances with up to 1000 nodes.

# Multistage multiscale facility location and expansion under uncertainty [*]

Laureano F. Escudero,[1] and Juan F. Monge[2]

[1] *Area of Statistics and Operations Research, Universidad Rey Juan Carlos, URJC, c/Tulipán, 28933 Móstoles (Madrid), Spain,*  laureano.escudero@urjc.es

[2] *Center of Operations Research, Universidad Miguel Hernández, UMH, Av. de la Universidad, 03202 Elche (Alicante), Spain,*  monge@umh.es

This work focuses on the development of a stochastic mixed integer linear optimization (traditionally, named MILP) modeling framework and a matheuristic approach for solving the multistage multiscale multiproduct facility location network expansion planning problem under uncertainty. Two types of time scaling are considered, namely, a long one and the other scale whose timing is much shorter. Then, two types of decisions are to be considered, viz., the strategic and the operational ones. The strategic decisions are the selection of facility locations in a network as well as the related initial capacity dimensioning and expansion along a time horizon. The operational decisions are the raw material supplying, the flow traffic through the available facility network, product manufacturing in some facilities and its distribution for demand satisfaction in some other available facilities at the minimum cost. Two types of uncertain parameters are also considered, namely, strategic and operational ones. It is assumed that the strategic uncertainty is stagewise-dependent, being captured by a finite set of scenarios that are represented in Hamiltonian paths from the first stage to the last one along the nodes in a multistage scenario tree. The operational uncertainty is stage-dependent, being captured by another type of a finite set of scenarios; the modeling scheme considers a set of two-stage trees, each one rooted at a node in the strategic multistage scenario tree. The goal is to minimize the expected total cost in the scenarios. Some strategic variables are binary and others are integer; they are the state variables linking a node with its successor ones. In any case, those variables are modelled by considering the step variable modeling object approach. It is tighter than its impulse counterpart one and, on the other hand, it implies that a node is only linked with its immediate successor ones, a feature that some decomposition algorithms for problem solving can take benefit from. By using the special structure of the location problems among others, the time-consistent risk averse measure Expected Conditional second order Stochastic Dominance is considered. Given the intrinsic problem's difficulty and the huge instances' dimensions (due to the network size of realistic instances as well as the cardinality of the strategic scenario tree and operational ones), it is unrealistic to seek an optimal solution. The matheuristic algorithm SFR3 is considered, it stands for Scenario variables Fixing and constraints and variables' integrality iteratively Randomizing Relaxation Reduction. It obtains a (hopefully, good) feasible solution in reasonable time and a lower bound of the optimal solution value to assess the solution quality. The performance of the overall approach is computationally assessed.

[*]Research partially supported by the Spanish Ministry of Science and Innovation through projects PID2019-105952GB-I00 and RTI2018-094269-B-I00

# A column-and-row generation algorithm for allocating airport slots

Paula Fermín Cueto,[1] Sergio García,[1] and Miguel F. Anjos[1]

[1]*School of Mathematics, The University of Edinburgh, The King's Buildings, Edinburgh, United Kingdom,*   paula.fermin@ed.ac.uk

Air transport demand often exceeds capacity at congested airports. For this reason, airlines need to be granted permission to use airport infrastructure. They must submit a list of regular flights that they wish to operate over a five to seven-month period and a designated coordinator is responsible for allocating the available airport slots, which represent the permission to operate a flight at a specific date and time. From an optimisation perspective, this problem is a special class of Resource Constrained Project Scheduling Problem (RCPSP) [1] where the objective is to minimise the difference between the allocated and requested flight times subject to airport capacity constraints and other operational restrictions.

Most studies on this topic focus on developing fast heuristics and complex models that capture the needs and particularities of various stakeholders [1, 3, 4]. It has been claimed that exact methods cannot cope with the size and complexity of real-world problems [5].

In this work we show that it is possible to find optimal solutions for large instances quickly and with modest memory requirements. We develop a column-and-row generation algorithm that uses the same principle as the *Zebra* algorithm for the *p*-median problem of García et al. [2]. Our algorithm capitalises on two interesting properties of airport slot allocation problems:

1. We have a good intuition about the optimal solutions, as it is known that, in real-world instances, the great majority of flights can be allocated to their requested time.

2. Most flights are regular services and airport capacity limits are typically constant throughout the season. This introduces an element of periodicity in the problem that results in a great number of identical or dominated capacity constraints.

We show the effectiveness of this algorithm using real-world data provided by Airport Coordination Limited (ACL) from the most congested airports in the United Kingdom.

# How to invest to expand a firm: a new model and resolution methods[*]

José Fernández,[1] Boglárka G.-Tóth,[2] and Laura Anton-Sanchez[3]

[1]*Dpt. Statistics and Operations Research, University of Murcia, Spain,*  josefdez@um.es

[2]*Dpt. Computational Optimization, University of Szeged, Hungary,*  boglarka@inf.szte.hu

[3]*Dpt. Statistics, Mathematics and Computer Science, Miguel Hernández University, Spain,*
l.anton@umh.es

When locating a new facility in a competitive environment, both the location and the quality of the facility need to be determined jointly and carefully in order to maximize the profit obtained by the locating chain. This fact has been highlighted in [1–3] among other papers.

However, when a chain has to decide how to invest in a given geographical region, it may also invest part of its budget in modifying the quality of other existing chain-owned centers (in case they exist) up or down, or even in closing some of those centers in order to allocate the budget devoted to those facilities to other chain-owned facilities or to the new one (in case the chain finally decides to open it). In this paper, we extend the single facility location and design problem introduced in [1] to accommodate these possibilities as well. The possibility of changing the quality of the existing chain-owned facilities makes the problem closer to the reality, but also harder to solve.

A mixed integer nonlinear programming formulation is proposed to model this new problem. Both an exact interval branch-and-bound method and an ad-hoc heuristic are proposed to solve the model. Some computational results are reported which show that both methods are able to solve this MINLP problem within a reasonable time and with good accuracy. According to the results, small variations in the available budget may produce very different results.

---

[*]Research supported by Fundación Séneca through project 20817/PI/18

# An exact method for the two-stage multi-period vehicle routing problem with depot location

Ivona Gjeroska,[1] and Sergio García [2]

[1] *School of Mathematics, University of Edinburgh,*   i.gjeroska@sms.ed.ac.uk

[2] *School of Mathematics, University of Edinburgh,*   sergio.garcia-quiles@ed.ac.uk

## 1.     Problem description

We introduce a vehicle routing problem (VRP) variation motivated by a company with a large distribution network in a city in Ecuador. The company sells a product of negligible size, works with roughly 1600 small retailers (customers), and has two types of "vehicles": sellers and trucks. Each customer is visited exactly once per week by a seller who takes the order. On the following working day, the same customer is visited by a truck that delivers their order. At the start of each day, the sellers meet at a meeting point that can vary. At the end of their working day they return to the depot. The trucks start and finish their routes at the depot. The goal is to find the optimal starting point for each working day for the sellers and create routes for the trucks and sellers that minimise the total travelling cost, such that the workload is balanced.

## 2.     Contribution

We introduced a multi-period VRP with depot location that consists of two stages that need to be solved simultaneously and a planning horizon with multiple periods with the addition of depot location. We provide two different mathematical models that fully describe the problem - one being a compact formulation with an adjustment to the degree constraints to serve the variation of the starting point, and one that is an adaptation of the standard capacitated VRP (CVRP). The latter comes in handy when applying existing algorithms for the CVRP. We introduce a tailor-made generalisation of the well known 2-matching and comb inequalities that are valid for this problem. These inequalities were first introduced and proven to be very efficient for the travelling salesman problem (TSP) [1], and later adapted for the CVRP [2]. Finally, we separate these inequalities using an adequate procedure that exploits the graph structure and uses cut-nodes and blocks as main tools. The idea comes from [3] who first used block structures to identify handles of violated 2-matching inequalities. We present results generated using a branch-and-cut algorithm, with and without the addition of the generalised comb inequalities. The test instances are subsets of the original set of customers provided by the company, created accounting for the geographical location and distribution of the original data set, preserving its main properties. For many of the instances the addition of the new valid inequalities proves to be beneficial: after two hours, it either results in a better solution or in a significantly lower number of nodes in the branch and bound tree. The latter is useful for larger instances, as the problem is less likely to run out of memory thus improving the chances to find a feasible solution.

# Formulations for the Capacitated Dispersion Problem[*]

Mercedes Landete,[1] Juanjo Peiró,[2] and Hande Yaman[3]

[1]*Universidad Miguel Hernández, Elche, Spain,*  landete@umh.es

[2]*Universitat de València, Burjassot (València), Spain,*  juanjo.peiro@uv.es

[3]*KU Leuven, Leuven, Belgium,*  hande.yaman@kuleuven.be

We study the capacitated dispersion problem (see [1]) in which, given a set $V$ of $n$ nodes (facilities), a positive capacity $c_i$ for each node $i$, a nonnegative distance $d_{ij}$ between any pair of distinct nodes $i$ and $j$, and a positive demand $B$ to cover, we would like to find a subset $V'$ of nodes such that the sum of capacities of nodes in this subset is large enough to cover the demand, i.e., $\sum_{i \in V'} c_i \geq B$ and the nodes in $V'$ are as distant as possible from one another, i.e., $\min_{i,j \in V': i \neq j} d_{ij}$, is maximum.

This problem arises, for instance, when we would like to locate facilities to cover the demand for a service, and we would like the facilities to be as distant as possible to decrease the risk of damage from accidents at other facilities.

In this talk we focus on several mathematical formulations for the problem in different spaces using variables associated with nodes, edges and costs. These formulations are then strengthened with families of valid inequalities and variable fixing procedures.

Several sets of computational experiments are conducted to illustrate the usefulness of the findings, as well as the aptness of the formulations for different types of instances.

---

[*]Research partially supported by the Spanish Ministry of Science and Innovation through projects PGC2018-099428-B-100 and RED2018-102363-T.

# Capacitated Close Enough Facility Location [*]

Alejandro Moya-Martínez,[1] Mercedes Landete,[1] Juan Franciso Monge[1] and Sergio García [2]

[1]*Centro de Investigación Operativa, Universidad Miguel Hernández, Elche,*  a.moya@umh.es, landete@umh.es,monge@.umh.es

[2]*School of Mathematics, University of Edinburgh,*  sergio.garcia-quiles@ed.ac.uk

Nowadays, companies face a continuously inreasing need of delivering goods. Mathematical programming models, and, in particular, location models can be used to improve these logistic activities. We find in the literature, various types of problems, to start with p-median problem to problems with cooperation between customers. In this paper, we present the Capacity - Close Enough Facility Location Problem (C-CEFLP). C-CEFLP is the problem of deciding where to locate $p$ facilities among the finite set of candidates, and where to locate $t$ pickup points close enough to customers. These pick up points have a limited capacity. Furthermore, it will be presented how this problem restricted to a graph behaves and a new column generation algorithm will be presented to solve it. We show that, the problem when the movements are restricted to a graph is computationally easier than the case without restrictions on the movements of the clients. A broad computational experiment is reported, and the performance of the heuristic approach is computationally assessed.

# References

[*]This research was funded by the Spanish Ministry of Science and Innovation and the State Research Agency under grant PID2019-105952GB-I00/AEI/10.13039/ 50110 0 011033, by the Spanish Ministry of Science and Innovation and the European Regional Development Fund under grant PGC2018-099428-B-100 and by the Spanish Ministry of Science and Innovation under the project RED2018-102363-T

# Emergency Vehicles Location:
# the importance of including the dispatching problem[*]

José Nelas[1,3] and Joana Dias[1,2]

[1]*University of Coimbra, Faculty of Economics, Av. Dias da Silva, 165, 3004-512 Coimbra, Portugal*

[2]*INESC Coimbra, University of Coimbra, Rua Sílvio Lima, Pólo II,3030-290 Coimbra Portugal*,  joana@fe.uc.pt

[3]*Centro Hospitalar e Universitário de Coimbra - Hospital Pediátrico, R. Dr. Afonso Romão, 3000-602 Coimbra Portugal*,  eunelas@gmail.com

Proper location of emergency vehicles is crucial to assure that assistance arrives on time where it is needed. The location of emergency vehicles is a strategic decision that highly influences the performance of the dispatching decisions: the choice of the vehicle that should be sent to an emergency episode. Although the dispatching decisions are operational decisions, they should be taken into account when deciding where to locate the vehicles, since dispatching decisions clearly influence the availability of resources. In this presentation, an emergency vehicle location model that explicitly considers resource availability by including dispatching decisions will be presented. Some results considering a real case study will also be shown.

## 1.      Introduction

The location of emergency vehicles represents an active research area, and many mathematical models and algorithmic approaches have been devel-oped, that consider different situations and that rely on different assumptions. The model to be presented has some distinguishing features that close the gap between model representation and the real situation of vehicle emergency management. This model explicitly considers, simultaneously, the existence of different vehicle types, capable of assuring different levels of care, and the possibility of one vehicle being substituted by another vehicle or set of vehicles that are equivalent in terms of the level of care they can provide. It is assumed that there is a set of potential and predetermined locations where the emergency vehicles can be located. The emergency episodes can occur anywhere within a predefined geographical area. Moreover, it is also possible to explicitly consider the evolution of the emergency episode, by assuming that one episode can have different stages and establishing different resource needs for these different stages depending on the evolution of the victims' health conditions. The model also represents the assumption that it is better to have some assistance arriving, even if it is not the most suitable one, than not having any assistance at all.

[*]This study has been funded by national funds, through FCT, the Portuguese Science Foundation, under project UIDB/00308/2020 and with the collaboration of Coimbra Pediatric Hospital – Coimbra Hospital and University Centre, and INEM.

## 2.     Real Case Study

The case study considers the emergency episodes that occurred in 2017, in the district of Coimbra, Portugal. All the data was totally anonymized and provided by INEM. In this civil year, a total of 50732 emergency episodes occurred, requiring 60343 vehicles' dispatches. It is possible to conclude that including the dispatching decisions significantly influences the location decisions. The solution calculated is able to achieve a better coverage of the geographic area considered than the current solution.

# Selective collection routes of urban solid waste by means of multi-compartment vehicles[*]

Ramón Piedra-de-la-Cuadra,[1] Juan A. Mesa,[2] Francisco A. Ortega[3] and Guido Marseglia [4]

[1] *Departamento Matemática Aplicada I, Universidad de Sevilla Spain,*  rpiedra@us.es

[2]*Departamento Matemática Aplicada II, Universidad de Sevilla, Spain,*  jmesa@us.es

[3]*Departamento Matemática Aplicada I, Universidad de Sevilla, Spain*  riejos@us.es

[4]*Departamento de Matemática Aplicada I, Universidad de Sevilla, Spain,*  marseglia@us.es

The rapid and constant increase in urban population has led to a drastic rise in urban solid waste production with worrying consequences for the environment and society. In many cities, an efficient waste management combined with a suitable design of vehicle routes (VR) can lead to benefits in the environmental, economic, and social impacts.

In recent years, the growth in urban population density has implied a major rise in the production of various kinds of Municipal Solid Waste (MSW), whose management includes several functional phases, such as waste generation, storage, collection, transportation, processing, recycling, and disposal in a suitable landfill. As a consequence, administrations, such as municipalities, have defined suitable waste collection areas to obtain efficiency and low environmental impact.

In Spain, the so-called eco-points are large waste containers with watertight sections to separate the collection of a wide variety of items. The management of eco-points gives rise to several problems that can be formulated analytically. The location and number of eco-point containers, the determination of the fleet size for picking up the collected waste, and the design of itineraries are all intertwined, but present computationally difficult problems, and therefore must be solved in a sequential way.

---

[*]Research partially supported by the Spanish Ministry of Science and Innovation through project RED2018-102363-T, Ministerio de Investigación (Spain)/FEDER under grant PID2019-533 106205GB-I00, and Operational Programme FEDER-Andalucía under grant US-1381656.

The mathematical optimization model formulated for this purpose has been identified as a combined version of BP problem and the VR problem, whose computational complexity motivates the use of heuristics to face large real-life scenarios. Following that recommendation, a greedy algorithm has been developed to solve the proposed mathematical programming model. Two strategies have been identified for designing the configurations of the mobile multi-block containers that will visit the demand nodes. The results obtained from the numerical simulations show the validation of the proposed methodology carried out for the Sioux Falls network benchmark and the specific real case study.

# Multi-Depot VRP with Vehicle Interchanges: Heuristic solution [*]

Victoria Rebillas-Loredo[1], Maria Albareda-Sambola[1], Juan A. Díaz[2], and Dolores E. Luna-Reyes[2]

[1] *Departament d'Estadística i Investigació Operativa, Universitat Politècnica de Catalunya, Spain*,   {victoria.rebillas, maria.albareda}@upc.edu

[2] *Department of Actuary, Physics and Mathematics, Universidad de las Américas Puebla (UDLAP), Mexico*   {juana.diaz, dolorese.luna}@udlap.mx

## 1.     Introduction

The classical variants of the Vehicle Routing Problem (VRP) aim at defining vehicle routes that, starting and ending at a given depot are able to provide service to a set of geographically scattered customers in the fastest/cheapest possible way. A natural extension of these problems consider settings where more than one depot is available. In those cases, each route is usually forced to finish at the same route where it started.

Among the most typical constraints considered in this type of problems we find the vehicle capacities, that limit the amount of customer demands that can be collected by each vehicle, and the limits on the total driving distance/time that can be assigned to each driver. Depending on the locations of the customers and on the distribution pattern of their demands, these constraints can cause optimal routes to underuse some of the resources; *i.e.* we can find short routes where the vehicle capacity constraints are binding, together with long routes where vehicle capacities are far from being completely used.

[*]Research partially supported by the Spanish Ministry of Science and Innovation through project RED2018-102363-T

The Multi-Depot Vehicle Routing Problem with Driver Interchanges (MD-VRPDI) explores the following strategy to overcome this undesired behavior:

Vehicle routes are decoupled from driver routes in a way that only driver routes are forced to start and finish at the same depot, while a vehicle can finish its route at a depot different from its starting point. To do so, several interchange points are set beforehand where two drivers can meet and interchange their vehicles. The reader is referred to [1] for furhter details.

This introduces additional flexibility to the model, allowing for a better usage of the available resources. However, by allowing these interchanges the complexity of the problem is heavily increased, since the interchanges require route synchronization.

# 2. Proposed heuristic

Even if, in the studied problem, the vehicle and driver routes are decoupled, for any feasible solution, the vehicle routes completely determine the driver routes. Therefore, solutions can be built by just focusing on the vehicle routes, and imposing feasibility of the resulting driver routes.

With this in mind, we observe that any feasible vehicle route can be obtained by combining two open vehicle routes. This combination must be made through an interchange point if the two open routes do not share the same depot, and without an interchange point if they both use the same depot. Moreover, if an interchange point is used, it must be used by an even number of combinations, yielding feasible driver routes.

The proposed heuristic applies different strategies to build promising open routes. Then, the above ideas are used to build a mathematical program that is used to iteratively combine open routes to build feasible solutions.

The performance of the heuristic is tested on a set of computational experiments.

# Optimizing COVID-19 Test and Vaccine distributions [*]

José Luis Sainz-Pardo,[1] José Valero[2]

[1]*Centro de Investigación Operativa (Universidad Miguel Hernández de Elche), Spain,* jose.sainz-pardo@umh.es

[2]*Centro de Investigación Operativa (Universidad Miguel Hernández de Elche), Spain,* jvalero@umh.es

The experience of Singapore and South Korea makes it clear that under certain circumstances massive testing is an effective way for containing the advance of the COVID-19. We propose a modified SEIR model which takes into account tracing and massive testing. After that, we introduce a heuristic approach in order to minimize the COVID-19 spreading by planning effective test distributions among the populations of a region over a period of time.

In a similar way, we propose a modified SEIR model which takes into account the effect of vaccination. The criteria to assess public health policies are fundamental to distribute vaccines in an effective way in order to avoid as many infections and deaths as possible. Usually these policies are focused on determining socio-demographic groups of people and establishing a vaccination order among these groups. This work also focuses on optimizing, from the proposed SEIR model, the way of distributing vaccines among the different populations of a region for a period of time once established the priority socio-demographic groups.

[*]Research partially supported by Generalitat Valenciana (Spain), project 2020/NAC/00022; Spanish Ministry of Science, Innovation and Universities, projects PGC2018-099428-B-I00 and PGC2018-096540-B-I00

# 1.    Test and vaccination models

On the one hand, we develop two SEIR models to forecast the evolution of the pandemic. SEIR models are compartmental models based on differential equations. Acronym SEIR is due to the fact that these models represent the evolution of Susceptible, Exposed, Infected and Recovered cases. The first introduced model, which is employed to optimize the test distribution, takes into account the impact of COVID testing and isolated cases. The second model, which is employed to optimize the vaccine distribution, takes into account the impact of vaccinated people.

# 2.    Parameters estimation and planning distribution

We use Differential Evolution technique for estimating the parameters. Regarding the optimal planning distribution, this is obtained from gain matrices computed from the SEIR models.

# 3.    Simulations and computational experience

For the computational test distribution experience, it has been simulated the pandemic spreading in New York counties from the first of April to the first of July of 2020. Table 1 shows the infections, the savings and the advantages in terms of infections to distribute several COVID test quantities both by homogeneous distribution and by the proposed approach.

Regarding the vaccine distribution we have reproduced the pandemic spread in the Spanish region called Valencian Community during the period from 1st of Juny to 31th of December of 2020. Figure 1 shows the infections by random vaccine distribution versus the proposed approach.

| # Tests | F. | H.Inf. | Ap.Inf. | H.Saving | Ap.Saving | Adv. |
|---|---|---|---|---|---|---|
| 10,000 | 1 | 3,365,783 | 3,365,709 | 34 | 108 | 74 |
| 10,000 | 3 | 3,365,716 | 3,365,494 | 101 | 323 | 222 |
| 10,000 | 9 | 3,365,514 | 3,364,851 | 303 | 966 | 663 |
| 50,000 | 1 | 3,365,666 | 3,365,280 | 151 | 537 | 386 |
| 50,000 | 3 | 3,365,365 | 3,364,209 | 452 | 1,608 | 1,156 |
| 50,000 | 9 | 3,364,462 | 3,361,026 | 1,355 | 4,791 | 3,436 |
| 100,000 | 1 | 3,365,519 | 3,364,744 | 298 | 1,073 | 775 |
| 100,000 | 3 | 3,364,924 | 3,362,612 | 893 | 3,205 | 2,312 |
| 100,000 | 9 | 3,363,144 | 3,357,124 | 2,673 | 8,693 | 6,020 |
| 500,000 | 1 | 3,364,346 | 3,360,256 | 1,471 | 5,561 | 4,090 |
| 500,000 | 3 | 3,361,416 | 3,354,512 | 4,401 | 11,305 | 6,904 |
| 500,000 | 9 | 3,352,700 | 3,341,480 | 13,117 | 24,337 | 1,1220 |
| 1,000,000 | 1 | 3,362,884 | 3,355,417 | 2,933 | 10,400 | 7,467 |
| 1,000,000 | 3 | 3,357,058 | 3,344,645 | 8,759 | 21,172 | 12,413 |
| 1,000,000 | 9 | 3,339,880 | 3,316,234 | 25,937 | 49,583 | 23,646 |

*Table 1.* Number of infected cases and saved infections with 10% tests per day limitation

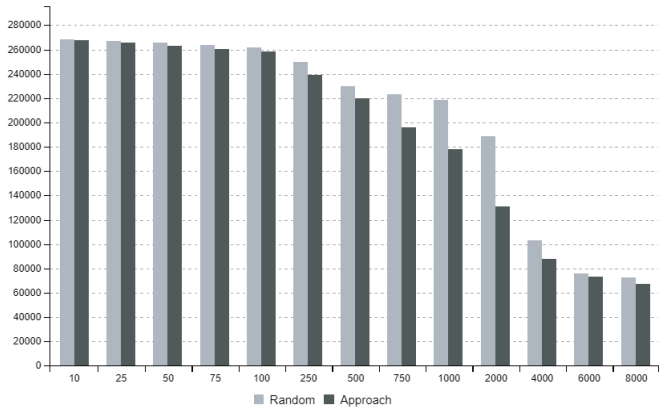

*Figure 1.* Infected cases by vaccine distribution methods

# Locating a rectangle in the sky to get the best observation[*]

Juan-José Salazar-González[1]

[1]*Departamento de Matemáticas, Universidad de La Laguna, Spain,*   jjsalaza@ull.edu.es

This paper concerns a new optimization problem arising in the management of a multi-object spectrometer with a configurable slit unit. The field of view of the spectrograph is divided into contiguous and parallel spatial bands, each one associated with two opposite sliding metal bars that can be positioned to observe one astronomical object. Thus several objects can be analyzed simultaneously within a configuration of the bars called a mask. Due to the high demand from astronomers, pointing the spectrograph's field of view to the sky, rotating it, and selecting the objects to conform a mask is a crucial optimization problem for the efficient use of the spectrometer. The paper describes this problem, presents a Mixed Integer Linear Programming formulation for the case where the rotation angle is fixed, presents a non-convex formulation for the case where the rotation angle is unfixed, describes a heuristic approach for the general problem, and discusses computational results on real-world and randomly-generated instances.

The combinatorial problem is related to locating a rectangle on a plane. It is an interesting problem in the Computational Geometry community, where there are many articles dealing with enclosing subsets of points with all kinds of geometric elements. Given a finite planar point set, the *enclosing problem* is to find the smallest geometrical element of a given type and arbitrary orientation that encloses all the n points. A kind of dual variant of the enclosing problem is finding the translation and orientation for a geometrical element of a given size to maximize the number of enclosed points.

[*]Research partially supported by the Spanish Ministry of Science and Innovation through project PID2019-104928RB-I00 (MINECO/FEDER, UE)

The full paper was recently accepted for publication in the journal Omega: https://doi.org/10.1016/j.omega.2021.102392 . At IWOLOCA we will summarize the main findings and will mention open questions.

# Multiple Allocation P-Hub Location Problem explicitly considering Users' preferences [*]

Nicolás A. Zerega,[1] Armin Lüer-Villagra,[2]

[1]*Department of Statistics and Operations Research, Universidad de Cádiz, Facultad de Ciencias, 11510 Puerto Real, Cádiz, Spain,*  nicolas.zerega@uca.es

[2]*Department of Engineering Sciences, Universidad Andres Bello, Antonio Varas 880, Piso 6, Santiago, Chile,*  armin.luer@unab.cl

Hub location problems (HLPs) are a well-known family of problems within General Network Design, which combine location and design decisions [1].

Hubs are facilities in which flow gets collected, bunched and then distributed to different nodes inside the network at reduced unitary costs thanks to the presence of economies of scale.

This type of problems have been widely studied in the last 30-40 years [2, 3].

In general, the entities that interact with the network are considered to be passive, i.e., their actions are based on the design decisions taken by the Network Manager. When the entities that interact with the network are humans, their preferences will not necessarily match with those of the Network Manager. For this reason, it is interesting to study how these individual decisions influence the performance of an existing network and, also, how they will influence the design of a new one.

This research develops the above idea by modelling users' decisions through deterministic utility functions [4]. Users try to maximize their utility when travelling from their Origin to their Destination nodes, meanwhile the Network Manager tries to maximize his/her profits.

We propose an extension of the Hub network design problem with profits (HNDPP) presented in Alibeyg et. al. 2016 [5] in which the users pref-erences are incorporated to the decision making process through the inclusion of Maximum Utility Constraints (MUC) [6]. This type of constraints implicitly model the users preferences and have been studied and applied in previous research [7,8].

[*]Research supported by the Chilean National Fund for Scientific and Technological Development (FONDECYT) through project 1200706

Current results show that the inclusion of the users' preferences has notable effects in the network design and performance, so considering them when making changes or designing a new network will provide a more realistic behavior.

# Author Index

The **International Workshop on Locational Analysis and Related Problems** will take place during January 31st – February 1st, 2022 in Elche (Spain).

It is organized by the Spanish Location Network and the Location Group GELOCA from the Spanish Society of Statistics and Operations Research (SEIO). The Spanish Location Network is a group of more than 140 researchers from several Spanish universities organized into 7 thematic groups. The Network has been funded by the Spanish Government since 2003. This edition of the conference is organized in collaboration with project PROMETEO/2021/063 funded by the Valencian government.